%% file: chemScience.tex
\documentclass[twoside,twocolumn]{article}
\usepackage[super,sort&compress,comma]{natbib} 
\usepackage[version=3]{mhchem}
\usepackage[left=1.5cm, right=1.5cm, top=1.785cm, bottom=2.0cm]{geometry}
\usepackage{balance}
\usepackage{booktabs,multirow,array}
\usepackage{times,mathptmx}
\usepackage{sectsty}
\usepackage{graphicx}
\usepackage{lastpage}
\usepackage[format=plain,justification=raggedright,singlelinecheck=false,font={stretch=1.125,small,sf},labelfont=bf,labelsep=space]{caption}
\usepackage{float}
\usepackage{fancyhdr}
\usepackage{fnpos}
\usepackage[english]{babel}
\usepackage{array}
\usepackage{droidsans}
\usepackage{charter}
\usepackage[T1]{fontenc}
\usepackage[usenames,dvipsnames]{xcolor}
\usepackage{setspace}
\usepackage[compact]{titlesec}
\usepackage{soul}
\usepackage[colorinlistoftodos,prependcaption,textsize=small]{todonotes}

\definecolor{cream}{RGB}{222,217,201}

\begin{document}

%\pagestyle{fancy}
%\thispagestyle{plain}
%\fancypagestyle{plain}{

%\newcommand{\alex}[1]{{\bf {\color{red} #1}}}
%\newcommand{\stalex}[1]{\iffalse #1 \fi}
%\newcommand{\stalex}[1]{\sout{#1}}

%%%PAGE SETUP - Please do not change any commands within this section%%%
\makeFNbottom
\makeatletter
\renewcommand\LARGE{\@setfontsize\LARGE{15pt}{17}}
\renewcommand\Large{\@setfontsize\Large{12pt}{14}}
\renewcommand\large{\@setfontsize\large{10pt}{12}}
\renewcommand\footnotesize{\@setfontsize\footnotesize{7pt}{10}}
\makeatother

\renewcommand{\thefootnote}{\fnsymbol{footnote}}
\renewcommand\footnoterule{\vspace*{1pt}% 
\color{cream}\hrule width 3.5in height 0.4pt \color{black}\vspace*{5pt}} 
\setcounter{secnumdepth}{5}

\makeatletter 
\renewcommand\@biblabel[1]{#1}            
\renewcommand\@makefntext[1]% 
{\noindent\makebox[0pt][r]{\@thefnmark\,}#1}
\makeatother 
\renewcommand{\figurename}{\small{Fig.}~}
\sectionfont{\sffamily\Large}
\subsectionfont{\normalsize}
\subsubsectionfont{\bf}
\setstretch{1.125} %In particular, please do not alter this line.
\setlength{\skip\footins}{0.8cm}
\setlength{\footnotesep}{0.25cm}
\setlength{\jot}{10pt}
\titlespacing*{\section}{0pt}{4pt}{4pt}
\titlespacing*{\subsection}{0pt}{15pt}{1pt}
%%%END OF PAGE SETUP%%%

%%%FIGURE SETUP - please do not change any commands within this section%%%
\makeatletter 
\newlength{\figrulesep} 
\setlength{\figrulesep}{0.5\textfloatsep} 

\newcommand{\topfigrule}{\vspace*{-1pt}% 
\noindent{\color{cream}\rule[-\figrulesep]{\columnwidth}{1.5pt}} }

\newcommand{\botfigrule}{\vspace*{-2pt}% 
\noindent{\color{cream}\rule[\figrulesep]{\columnwidth}{1.5pt}} }

\newcommand{\dblfigrule}{\vspace*{-1pt}% 
\noindent{\color{cream}\rule[-\figrulesep]{\textwidth}{1.5pt}} }

\newcommand{\figref}[1]{Fig.~\ref{#1}}

\newcommand{\rev}[1]{\textcolor{black}{#1}}
\newcommand{\mfs}[1]{\textcolor{green}{MFS: #1}}
\newcommand{\arr}[1]{\textcolor{red}{ARR: #1}}
\newcommand{\lsp}[1]{\textcolor{cyan}{LSP: #1}}
\newcommand{\pb}[1]{\textcolor{orange}{PB: #1}}

\makeatother
%%%END OF FIGURE SETUP%%%

%%%TITLE, AUTHORS AND ABSTRACT%%%
%\twocolumn[
 % \begin{@twocolumnfalse}
%\vspace{3cm}
%\sffamily
%\begin{tabular}{m{4.5cm} p{13.5cm} }

%\includegraphics{head_foot/DOI} & 

\noindent\LARGE{\textbf{Bias-dependent local structure of water molecule at a metallic interface}} \\
%\vspace{0.3cm} & \vspace{0.3cm} \\

\noindent\large{Luana S. Pedroza,\textit{$^{a}$} Pedro Brandimarte,\textit{$^{b}$} Alexandre Reily Rocha,\textit{$^{c}$} and M.-V. Fern\'andez-Serra\textit{$^{d}$}} \\

%\includegraphics{head_foot/dates} & 

%--------------------------------------------------------------------------------------------------

\noindent\normalsize{Understanding the local structure of water at the interfaces of metallic electrodes is a key problem in aqueous-based electrochemistry.
Nevertheless, a realistic simulation of such setup is challenging, particularly when the electrodes are maintained at different potentials.
To correctly compute the effect of an external bias potential applied to truly semi-infinite surfaces, we combine Density Functional Theory (DFT) and Non-Equilibrium Green's Functions (NEGF) methods.
This framework allows for the out-of-equilibrium calculation of forces and dynamics, and directly correlates to the chemical potential of the electrodes, which is the one introduced experimentally.
In this work, we apply this methodology to study the electronic properties and atomic forces of one water molecule at the interface of gold surface.
We find that the water molecule tends to align its dipole moment with the electric field, and it is either repelled or attracted to the metal depending on the sign and magnitude of the applied bias, in an asymmetric fashion.} \\

%\end{tabular}

% \end{@twocolumnfalse} \vspace{0.6cm}

%%%END OF TITLE, AUTHORS AND ABSTRACT%%%

%%%FONT SETUP - please do not change any commands within this section
\renewcommand*\rmdefault{bch}\normalfont\upshape
\rmfamily
\section*{}
\vspace{-1cm}

%%%FOOTNOTES%%%

\footnotetext{\textit{$^{a}$~ICTP South American Institute for Fundamental Research,
Instituto de F\'isica Te\'orica, Universidade Estadual Paulista, S\~ao Paulo SP 01140-070, Brazil. Centro de Ci\^encias Naturais e Humanas, Universidade Federal do ABC, Santo Andr\'e, S\~ao Paulo, Brazil 09210-170. E-mail: l.pedroza@ufabc.edu.br}}
\footnotetext{\textit{$^{b}$~Centro de F\'isica de Materiales, 20018 Donostia-San Sebasti\'an, Basque Country, Spain. Donostia International Physics Center, 20018 Donostia-San Sebasti\'an, Basque Country, Spain.}}

\footnotetext{\textit{$^{c}$~Instituto de F\'isica Te\'orica, Universidade Estadual Paulista, S\~ao Paulo, S\~ao Paulo SP 01140-070, Brazil.}}

\footnotetext{\textit{$^{d}$~Department of Physics and Astronomy,  Stony Brook University, Stony Brook, New York 11794-3800, USA.
Institute for Advanced Computational Sciences, Stony Brook University, Stony Brook, New York 11794-3800, USA.}}

%Please use \dag to cite the ESI in the main text of the article.
%If you article does not have ESI please remove the the \dag symbol from the title and the footnotetext below.
%\footnotetext{\dag~Electronic Supplementary Information (ESI) available: [details of any supplementary information available should be included here]. See DOI: 10.1039/b000000x/}
%additional addresses can be cited as above using the lower-case letters, c, d, e... If all authors are from the same address, no letter is required

%\footnotetext{\ddag~Additional footnotes to the title and authors can be included \emph{e.g.}\ `Present address:' or `These authors contributed equally to this work' as above using the symbols: \ddag, \textsection, and \P. Please place the appropriate symbol next to the author's name and include a \texttt{\textbackslash footnotetext} entry in the the correct place in the list.}

%%%END OF FOOTNOTES%%%

%%%MAIN TEXT%%%%

%==================================================================================================
\section{Introduction}

Following the need for new - and renewable - sources of energy worldwide, fuel cells using electrocatalysts can be thought of as a viable option \cite{Norskov09, Greeley12}.
The interface between a metal (electrode) and water in these systems is the electrochemical central point, since it is the region where charge transfer can take place.
A better understanding of the metal-water interface is also an essential requisite for predicting the correspondence between the macroscopic voltage and the microscopic interfacial charge distribution in electrochemical fuel cells.
This reactivity is governed by the explicit atomic and electronic structures built at the interface as a response to external conditions, such as an applied potential \cite{AHodgson09, Madley87, interfaceReview}.

The advance in  experimental techniques for studying surfaces in the last decades started to provide important results concerning the local structure of water at interfaces, revealing a bias-dependent behavior \cite{Toney94, SalmeronScience14, OsawaJPC96}.
Notably, in a more realistic system applied to catalysis, the metal will be at a given potential.  
From a theoretical perspective it makes the task of simulating this setup difficult \cite{JRossmeislSS15}.
In fact, an accurate calculation of the electrostatic potential at electrically biased metal-electrolyte interfaces is a challenge for {\it ab initio} simulations with periodic boundary conditions \cite{MKoper, MSprik}.

One possibility to simulate this electrochemical cell under an explicit bias is to account for the polarization of the metal by charging each atom on the electrode, enforcing a constant potential and using the image charges  method \cite{Siepmann-Sprik, DChandlerFD09, GVothJPCC12}.
Although they can provide interesting insights into the problem, the description is limited to the use of empirical models.
Neurock's studies of the water/metal interface in the presence of an applied potential \cite{Neurock, Neurock06} are among the first ones that have used Density Functional Theory (DFT) \cite{dft1,dft2} to address this problem.
The resulting electrode potential -- which is related to the Fermi energy of the system --  is compared to an internal reference potential by artificially inserting a vacuum layer into the center of the solution region.
To be able to fully compute all these energies when the system is charged, the water molecules in the center of the liquid layer are usually held fixed during the optimization of the charged systems.
More recently, N. Bonnet \emph{et al.} proposed a  methodology where an external potentiostat is added to the system \cite{SuginoPRL12}, following a previous work where the charges at the surface are controlled by including a medium with a given permittivity\cite{SuginoPRB2006}.
The idea in the later work is to use the potential energy of a fictitious system, akin to the fictitious mass in Car-Parrinello first-principles molecular dynamics. 
Therefore, current methodologies\cite{FilholPCCP2011} attempt to simulate the effect of finite bias at the metal by altering their charge (adding/subtracting electrons), whereas in experiments the potential is the quantity that is controlled.

The electrochemical cell can be thought of as two metallic electrodes which act as charge reservoirs, with the two metal plates separated by a solution (mostly composed by water).
This is an arrangement analogous to the one encountered in simulations of electronic transport: a central scattering region coupled to electrodes \cite{datta,transiesta,smeagol}.
Thus, we propose in this work, as an alternative, to use open boundary conditions by employing the non-equilibrium Green's function (NEGF) method combined with DFT to properly  compute the effect of an external bias potential applied to electrodes.
While standard DFT implementations are not suited to treat extended systems under an external bias, NEGF has been designed to treat out-of-equilibrium situations.
Most notably it allows for the inclusion of truly semi-infinite metallic electrodes, which set the correct chemical potential for the metal, and a clear reference potential.
Their combination has been developed over the past decade to describe current-voltage characteristics of nanoscopic systems \cite{smeagol,smeagol2,transiesta}.
It treats an open system under the influence of an external bias, and albeit dynamics - or forces - is typically ignored in such systems, it can be incorporated into the methodology.
In this work, we apply this framework to a system consisting of a single water molecule between two Au(111) surfaces, at different configurations and as a function of an external voltage.

%==================================================================================================
\section{Methodology}

%--------------------------------------------------------------------------------------------------
\subsection{General Methodology}

\begin{figure}[!htb]
\includegraphics[width=\columnwidth]{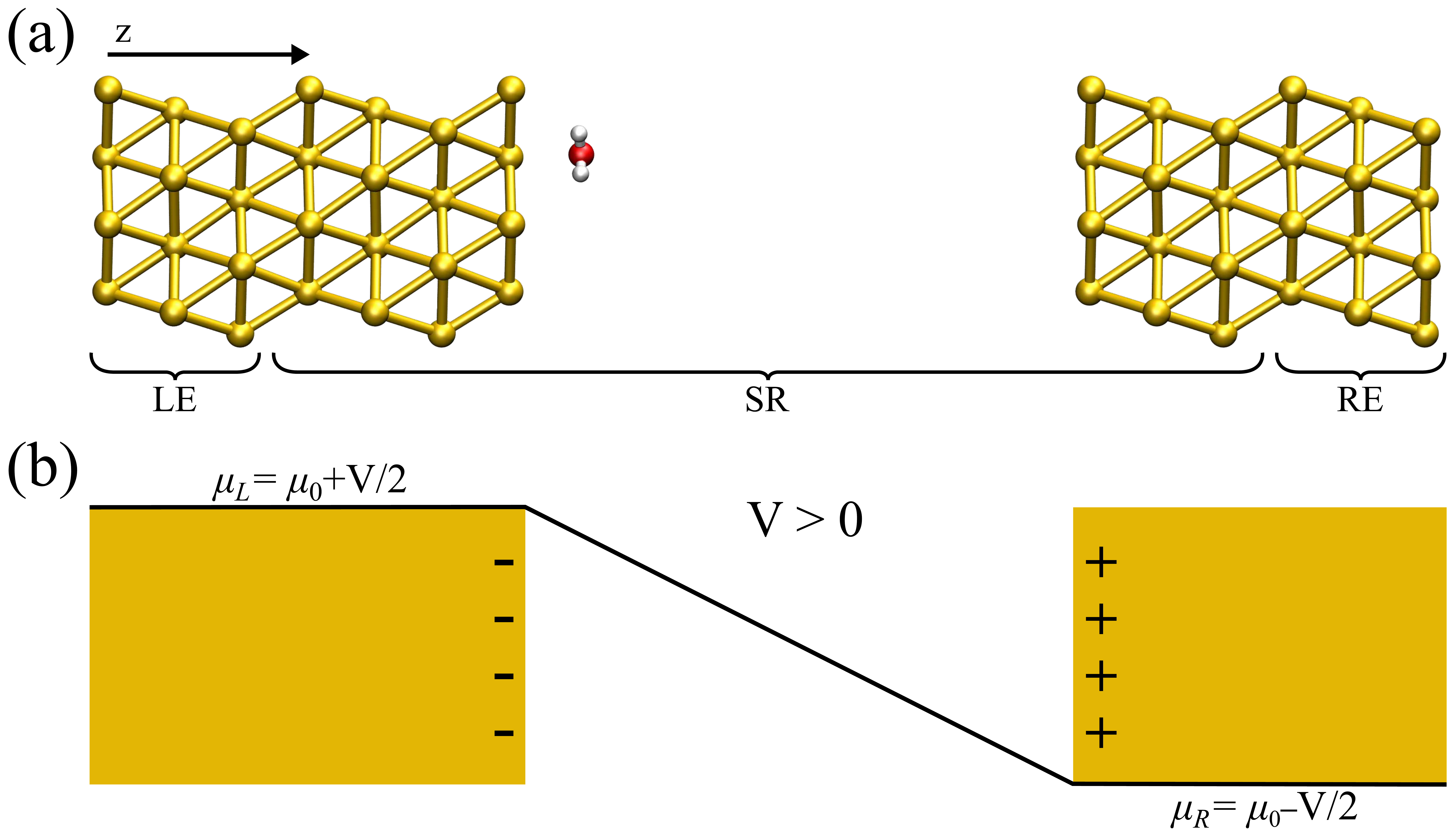}
\caption{(Color online) (a) Schematic view of the metal-water system used for the non-equilibrium calculations; the left (right) electrodes (LE/RE) and scattering region (SR) are indicated.
(b) Sketch of the effect of a positive bias potential on a parallel plate capacitor;
the corresponding charge accumulated in each plate as well as the bias ramp are shown.}
\label{Fig1-system}
\end{figure}

In what has become an usual approach in electronic transport calculations, the system is divided into three regions: electrodes (left and right, L/R) and scattering region (SR) \cite{caroli1971}.
We can frame our metal-water system into an analogous arrangement.
In this case, the interfacial region (metal-water-metal) represents the scattering region.
The first few layers of the metal at the interface need to be considered part of the scattering region as we require that the charge density at the edge of the SR resembles the one of the bulk metal.
Then, the electronic charge distributions in the electrodes, left and right, correspond to the bulk phases of the same material to a prescribed numerical accuracy.
A representation of the typical arrangement used in our calculations is shown in \figref{Fig1-system}(a). 

When a finite voltage is applied to the electrodes the problem becomes a non-equilibrium one.
The electrodes are then ascribed different chemical potentials and current, in principle, can flow.
The non-equilibrium Green's function formalism is a general formalism for calculating the properties of systems in out-of-equilibrium situations, and can be used to tackle our problem of the electrochemical cell.
In principle it can be used to address problems where inelastic effects are present, and most importantly, it goes far beyond electronic transport (including ballistic transport).

Within the NEGF approach, if the Hamiltonian can be cast in a bilinear form, the entire problem can be treated in terms of a single Green's function;
in our case the retarded Green's function for the scattering region,
\begin{eqnarray}
G^\mathrm{\left(E,V\right)} & = & lim_{\eta\rightarrow 0^+} \left[ \epsilon^+ S_\mathrm{SR} - H_\mathrm{SR}\left[n\right] \right. \notag \\
 & & \left.  -\Sigma_\mathrm{L}\left(E,V\right) - \Sigma_\mathrm{R}\left(E,V\right)\right]^{-1} \,,
\end{eqnarray}
where $\epsilon^+ = E + i\eta$, $S_\mathrm{SR}$ is the overlap matrix and $H_\mathrm{SR}\left[n\right]$ is a Hamiltonian which is a functional of the charge density, $n(\vec r)$. In this work the Hamiltonian is taken as the Kohn-Sham (KS) hamiltonian from DFT. The effects of the electrodes are introduced in the form of the self-energies $\Sigma_\mathrm{L/R}$, which are obtained by integrating out the degrees of freedom of the leads. As the electrodes are considered to be good metals, the effect of the bias on the left and the right electrodes corresponds to a rigid shift ($\pm V/2$) of the zero-bias self-energies, setting the boundary condition (illustrated in \figref{Fig1-system}(b) for a positive bias). This means that the self-energies can be obtained from a separate DFT calculation for the bulk metal, and need not to be updated self-consistently throughout the calculation. This approach also ensures that a clear reference potential is defined (the chemical $\mu_0$ of the bulk metal) as we assume the electrodes are charge reservoirs in thermodynamic equilibrium throughout the calculations.

Once the Green's function of the SR is calculated, all the observables of the system can be recomputed. In particular the density matrix is expressed as:
\begin{equation} \label{eqRhoNEQ}
D_{\mu\nu}=\int_{-\infty}^{\infty} dE [\rho^L_{\mu\nu} f (E - \mu_L) + \rho^R_{\mu\nu}f (E - \mu_R) ] \,,
\end{equation}
where $\mu$ and $\nu$ indexes run over the SR electronic states, $f (E) = 1/(1+e^{E/kT})$ is the Fermi distribution and $\mu_R$ and $\mu_L$ are the electrochemical potentials of the right and left electrodes ($\mu_{L/R}=\mu_0 \pm V/2$), that defines the bias: $V=\mu_L - \mu_R$.
Finally
\begin{equation}
\rho^{L/R}=G\left(E\right)\Gamma_{L/R}\left(E\right)G^\dagger\left(E\right)
\end{equation}
is the electrode spectral density matrix, obtained from the Green's function of the SR and the left (right) coupling matrices, $\Gamma_{L/R} = i\left[\Sigma_{L/R}\left(E\right)-\Sigma_{L/R}^\dagger\left(E\right)\right]$.
From the density matrix, the KS hamiltonian can be computed.
The procedure is then repeated until self consistency is achieved.

Within the ground state DFT framework, the computation of forces on the nuclei is theoretically well founded thanks to the Hellman-Feynman theorem\cite{HFtheorem1,HFtheorem2}, and the forces are obtained via the derivative of the total energy.
Using a set of localized basis functions the force is decomposed in two terms\cite{SanvitoForces}:
one that originates from the derivative of  the energy of the occupied eigenstates (band structure contribution) and a second one that contains the remaining contributions to the energy.
For an ion $I$, the former one is given by:
\begin{equation} \label{eqForce}
\vec F^{BS}_I=-\sum_{\mu \nu} D_{\mu \nu} \frac{\partial H_{\mu \nu}}{\partial \vec R_I} + \Omega_{\mu \nu} \frac{\partial S_{\mu \nu}}{\partial \vec R_I} ~,
\end{equation}
where $D_{\mu \nu}$ is the density matrix and $\Omega_{\mu \nu}$ is the energy density matrix ($\Omega_{\mu \nu}=\sum_i E_i f(E_i) \Psi_{i\mu}\Psi^*_{i\nu}$).
The situation is more complex out of equilibrium, where the Hellman-Feynman theorem does not apply\cite{DiVentra}.
Recently, it was shown that the forces can actually be obtained by the time derivative of the expectation value of the ionic momentum operators\cite{DiVentra,SanvitoForces}:
\begin{equation}
\vec F_I = \frac{\partial }{\partial t} \left \langle \Psi (t) \right| -i \hbar \frac{\partial}{\partial \vec R_I} \left| \Psi (t) \right\rangle \,.
\end{equation}
As it turns out, for steady-state problems, the final form for the force is equivalent to the equilibrium case,
\begin{equation}
\vec F_I = - \frac{\partial \langle \Psi | H| \Psi \rangle}{\partial \vec R_I} ~,
\end{equation}
which can be expressed by Eq. \ref{eqForce}, replacing the ground state density matrix and energy density matrix by the out-of-equilibrium ones, now obtained in terms of the retarded Green's function.

One important point that still remains is how well defined are the forces when current flows in the system \cite{Todorov,SanvitoCurrentForce,MBrandbyge}.
In our case, however, the gap in the scattering region -- the band gap of water -- is large enough ($\sim 8$ eV) to ensure that no current will flow through the arrangement.
In that sense, it is important to stress that current-induced forces will not be present in our problem.
This remains true for the simulation of ionic electrolytes, where ionic currents are expected to exist - and can be captured by this method - but no electronic currents.

\rev{Finally, one notices that the above methodology relies on the Kohn-Sham Hamiltonian being a good description of the single particle excitations for the system, as it is the case in different implementations of the NEGF formalism within DFT \cite{transiesta, smeagol}.
This leads to known pitfalls, which are associated to position of molecular energy levels and charge transfer between surface and molecule to name a few \cite{PhysRevLett.97.216405, PhysRevLett.106.187402}.
Most of these issues however, pertain to the approximations in the exchange-correlation functional, and  corrections in different forms can be readily incorporated into the formalism \cite{PhysRevLett.95.146402, PhysRevB.83.115108, PhysRevB.82.125426, PhysRevB.88.165112, C4NR04081C}.
Nonetheless, it is important to point out, that local and semi-local functionals tend to perform better for forces and structures compared to total energies and single particle energy levels.
}

The described methodology was implemented in the Smeagol code \cite{smeagol, smeagol2} which is bundled with Siesta \cite{siesta1, siesta2}.
In the same way that relaxation and \emph{ab initio} molecular dynamics can be performed within DFT, one can now use the code to do the same for out-of-equilibrium systems.

%--------------------------------------------------------------------------------------------------
\subsection{Details of calculations}

In this work, we have used two different gradient-dependent exchange-correlation (XC) functionals: PBE \cite{pbe} and vdW-DF$^{PBE}$, which includes van der Waals corrections (vdW).
The vdW-DF$^{PBE}$ is a modified version of the original vdW-DF functional \cite{drsll}, in which the revPBE local term was replaced by PBE \cite{JWangJCP11}.
The core electrons were described by norm-conserving pseudopotentials in the Troullier-Martins form \cite{pseudo1}.
A basis set of numerical atomic orbitals with double-$\zeta$ polarization was used to describe the valence electrons.
For both metal and water the basis set was variationally optimized and ensured that our results (Au lattice parameter, water-metal geometry) are in agreement with plane-wave calculations.

For the non-equilibrium calculations, each metal slab within the scattering region has 3 layers of (111) planes with 12 Au atoms on each plane, with size $10.29\times9.89$ \AA\ in the plane perpendicular to the transport direction.
This size is chosen because periodic boundary conditions are still applied in this plane, and it is necessary to minimize the interaction between periodic repetitions of the water molecules in the plane.
The water molecule is placed close to one metal surface (the one defined as left).
In order to minimize the interaction between the surfaces, the right and left side are 20 \AA\ apart.
The electrodes, connected to the scattering region, consist of 3 Au layers each (left and right).
\figref{Fig1-system} shows a schematic view of the system and its components.

%==================================================================================================
\section{Results}

\begin{figure}[!htb]
\includegraphics*[width=\columnwidth,keepaspectratio]{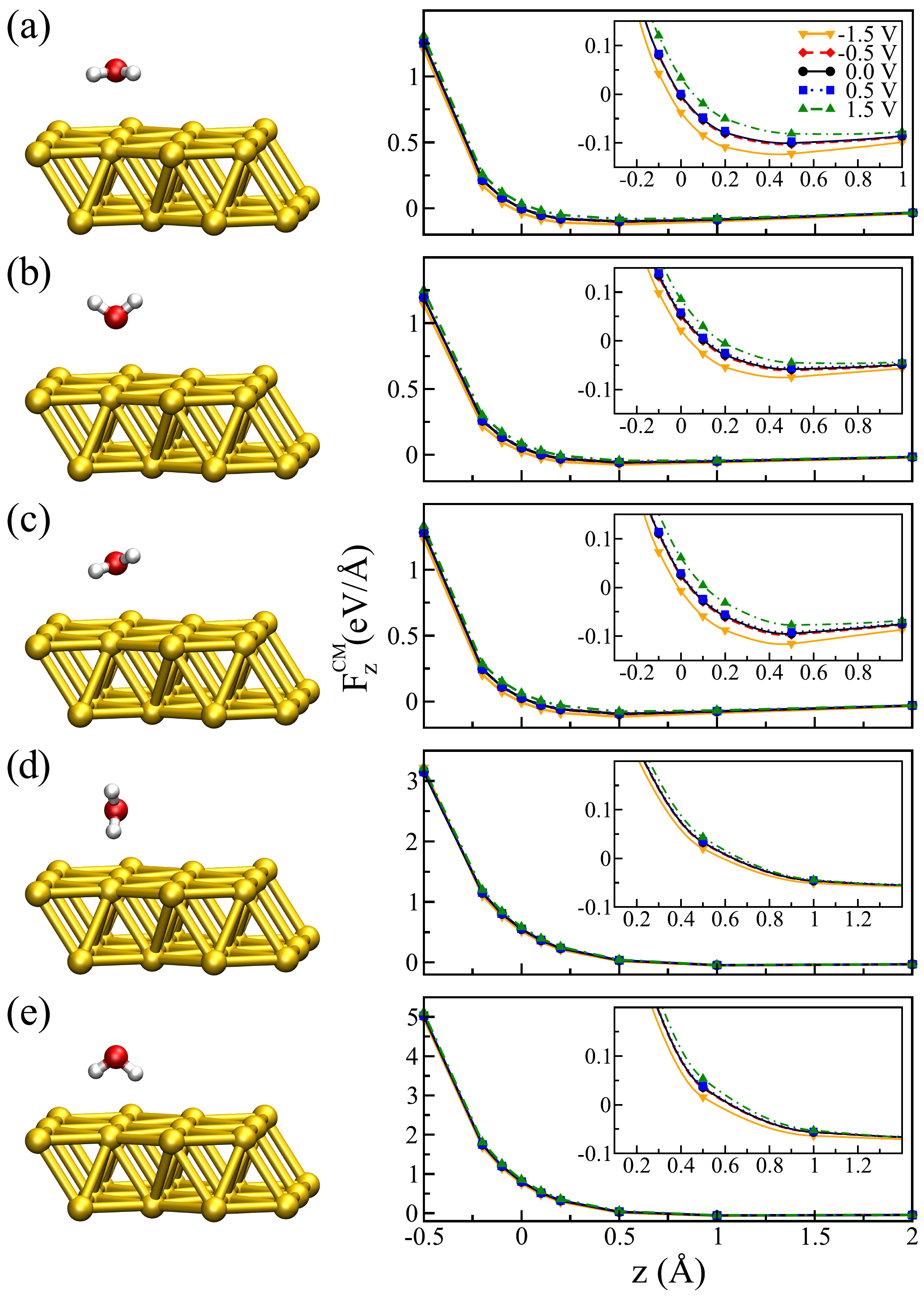}
\caption{(Color online) \rev{$z$-component} of the force at the center of mass \rev{(right panels)} as a function of the vertical displacement for different water configurations \rev{(left panels)}: (a) ``flat'', (b) \rev{``up''}, (c) ``\rev{flat-up}'', (d) ``\rev{perpendicular}'' and (f) ``\rev{down}''.
The vertical displacement is given with respect to the Au-O distance in the ground state. \rev{The insets show the regions in each graph for which $F_z^{CM}=0$.}}
\label{Fig3-FcmAll}
\end{figure}

Before applying a bias at the electrodes it is important to characterize the ground state configuration of the metal-water system.
This was \rev{initially} done using a (111) surface Au slab with 4 layers and a 2x2 in-plane supercell within the standard, periodic DFT formalism.
\rev{The relaxed water structures were then used as  starting configurations for the larger gold surfaces.}
All the atoms were allowed to move during the geometrical optimization, using the conjugate gradient algorithm and with a 0.005 eV/\AA\ tolerance criteria on the forces.
\rev{The final configurations were very similar to the ones used as a starting point.}
Our results show that the most stable configuration of one water molecule on top of the Au(111) surface is the so-called ``flat'' one (i.e. the molecule dipole moment is almost parallel to the surface plane), in agreement with previous reported calculations \cite{MichaelidesPRL03,Cicero}.
The molecular plane is slightly tilted with respect to the metal plane, with $\alpha=3^\circ$ ($\alpha$ being the angle between the molecular plane and the surface plane) and the distance between Au and O, $d_{Au-O}= 2.92$ \AA.
This structure is illustrated in \figref{Fig3-FcmAll}(a). We have also relaxed the metal-water-metal structure using the NEGF formalism at zero bias, \rev{obtaining $\alpha=6^\circ$ and 2.79 \AA\ for the Au-O distance.
The small differences are attributed to the effect of using a finite representation of Au surface in the standard DFT calculation.}

It is worth mentioning that the potential energy surface (PES) for this system is very flat in the region of the water on top of Au.
For instance, the energy difference between the configuration where the molecule is ``flat'' compared to the one where it has the hydrogens pointing down (towards the metal) is only $\sim 0.06$ eV.
Therefore, we also considered other four different configurations for the water molecule, corresponding to rigid rotations of the ground state structure \rev{(left panels of \figref{Fig3-FcmAll})}.
The geometries are labeled according to their orientation: ``up'' (hydrogens pointing away from the metal), ``down'' (hydrogens pointing towards the metal), ``perpendicular'' ($\alpha=90^\circ$ and $\theta=90^\circ$), and ``flat-up'' (one hydrogen higher than the flat configuration, with $\alpha=24^\circ$ and $\theta=84^\circ$).
The angle $\theta$ corresponds to the angle between the molecule dipole and surface normal.

In order to analyze the effect of different bias voltages (magnitude and sign) on the molecule as a function of the distance of the water molecules to the metal, we first performed non-equilibrium calculations for all water structures.
For each configuration, we started at the ground state Au-O distance ($z=2.79$ \AA, which corresponds to zero in the plots) \rev{at zero bias} and increased/decreased this distance by -0.5 to +2.0 \AA, and performed a single point calculation.
At each point the forces on the atoms were evaluated for a particular applied bias.
Since the water molecule is placed close to one metal surface, we observe that the potential in the water molecule follows closely that of the surface.
Therefore, the bias we are indicating in the plots corresponds to $V/2$, as it corresponds to the potential effectively acting on the molecule.

The results for the \rev{$z$-component} of the force on the center of mass of the molecule\rev{, $F_z^{CM}$} are shown in \figref{Fig3-FcmAll} for all orientations.
In general, we observed that low bias (-0.5 and +0.5 V) has a small effect on the forces, independently of the water configuration.
However, as we increase the applied bias we observe that the forces close to the minimum are modified.
In particular, this effect is more evident for the flat molecule (\figref{Fig3-FcmAll}(\rev{a})) and for configurations where the oxygen is facing the metal, due to strong interaction between the oxygen-$b_1$ orbital of the water molecule and the metal orbitals \cite{Adrien11, MichaelidesJCP09}.

\figref{Fig3-FcmAll} indicates that there is a tendency to modify the position of the minimum configuration when the bias is applied.
Moreover, this modification is dependent on the sign of the bias as it is asymmetric with respect to positive and negative values.
This behavior is similar to what is verified when an electric field is applied \cite{EfieldH2O, noteEfield}:
The molecule tends to get closer to the metal when the bias is negative and moves away from the metal with positive bias\rev{, evidenced by the position at which $F_z^{CM}$=0.} A similar trend can be observed for configurations ``up'' and ``flat-up'' shown in \figref{Fig3-FcmAll}(\rev{b-c}), respectively. For ``perpendicular'' and ``down'' configurations the molecule is essentially unbound (\figref{Fig3-FcmAll}(\rev{d-e}), respectively).

\begin{figure}[!htb]
\begin{center}
\includegraphics*[width=\columnwidth,keepaspectratio]{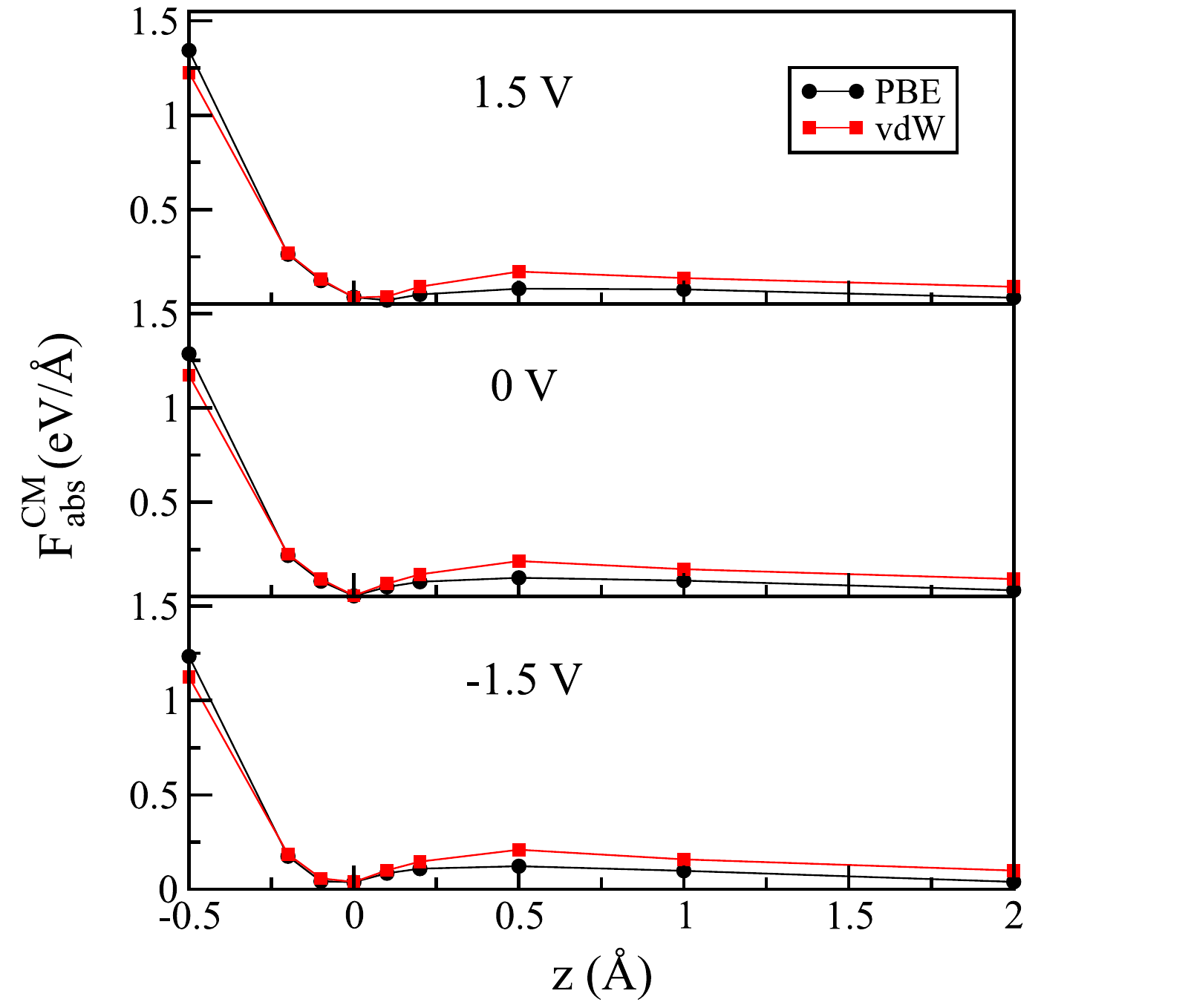}
\end{center}
\caption{(Color online) Magnitude of the force at the center of mass as a function of the vertical displacement for the ``flat'' configuration for PBE and vdW-DF$^{PBE}$ exchange-correlation functionals.
The vertical displacement is given with respect to the Au-O distance at ground state correspondent to each functional.}
\label{Fig4-vdw}
\end{figure}

A similar behavior was also observed when vdW corrections are included for the flat configuration, as shown in \figref{Fig4-vdw}.
In agreement to previous work \cite{LPedroza}, we note that the vdW functional does not change significantly the water-Au interaction.
We observe, however, that the barrier in all cases increases slightly for higher distances; this is true for zero as well as for finite (positive or negative) bias.
This means that, although the equilibrium position of the molecule does not depend on the choice of XC functional \rev{(specifically for Au-water systems)}, the restoration force is larger when we include vdW interaction.

\begin{figure}[!htb]
\begin{center}
\includegraphics*[width=0.9\columnwidth,keepaspectratio]{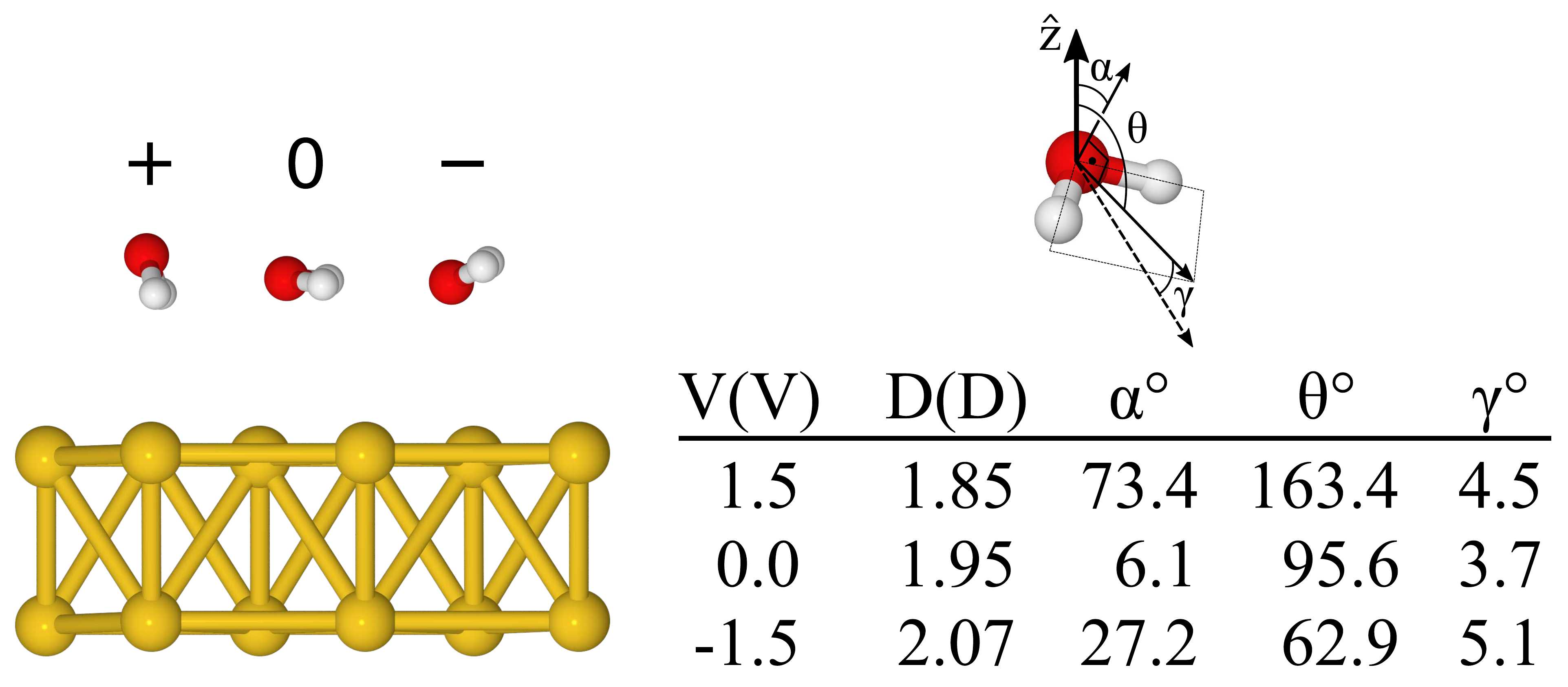}
\end{center}
\caption{(Color online) Relaxed configurations of the water molecule on the Au surface for different bias voltages ($+1.5$ V $\rightarrow +$ , 0 V, $-1.5$ V $\rightarrow -$), the corresponding angles,  \rev{and estimated dipole moment (details in Supplementary Information)}.
The angle $\alpha$ is the angle between the molecular plane and the surface plane, $\theta$ corresponds to the angle between the isolated molecule dipole and surface normal, \rev{and $\gamma$ is the angle between the isolated molecule dipole and the one calculated from the charge density of the combined Au+H$_2$O system}.}
\label{minima}
\end{figure}

\rev{Although, by rigidly shifting the position of the molecule along $z$, one can always find a position for which $F_z^{CM}$=0, that is not the case for all directions concomitantly (see Supplementary Information).
This is an indication that, as the absolute value of the bias increases, the orientation of the molecule tends to change as well.
Thus, i}n a second step, starting from the ``flat'' configuration, we allowed the atoms of the water molecule to move using the conjugate gradient algorithm with a bias applied to the system.
The minimum configuration obtained for -1.5 and +1.5 V are shown in \figref{minima}, where the geometry obtained for the zero bias case is also shown.
The asymmetric behaviour with respect to the bias sign is clearly observed.
The geometry for +1.5 V has the Hydrogen atoms pointing down and the Oxygen atom is 2.84 \AA\ far away from the metal.
In fact, this is the ``down'' configuration ($\alpha=90^\circ$ and $\theta=180^\circ$), presented in \figref{Fig3-FcmAll}(\rev{e}), which means that the positive bias leads to an unbound molecule.
On the other hand, when the negative bias is applied the Hydrogen atoms slightly move upwards ($\alpha=27^\circ$ and $\theta=63^\circ$) and the Oxygen gets closer to the metal ($d_{O-Au}=2.69$ \AA) when compared to the neutral case.
In essence, one notices that even a relatively small bias, can lead to significant structural changes on the metal-water interface.

\begin{figure}[!htb]
\includegraphics*[width=\columnwidth,keepaspectratio]{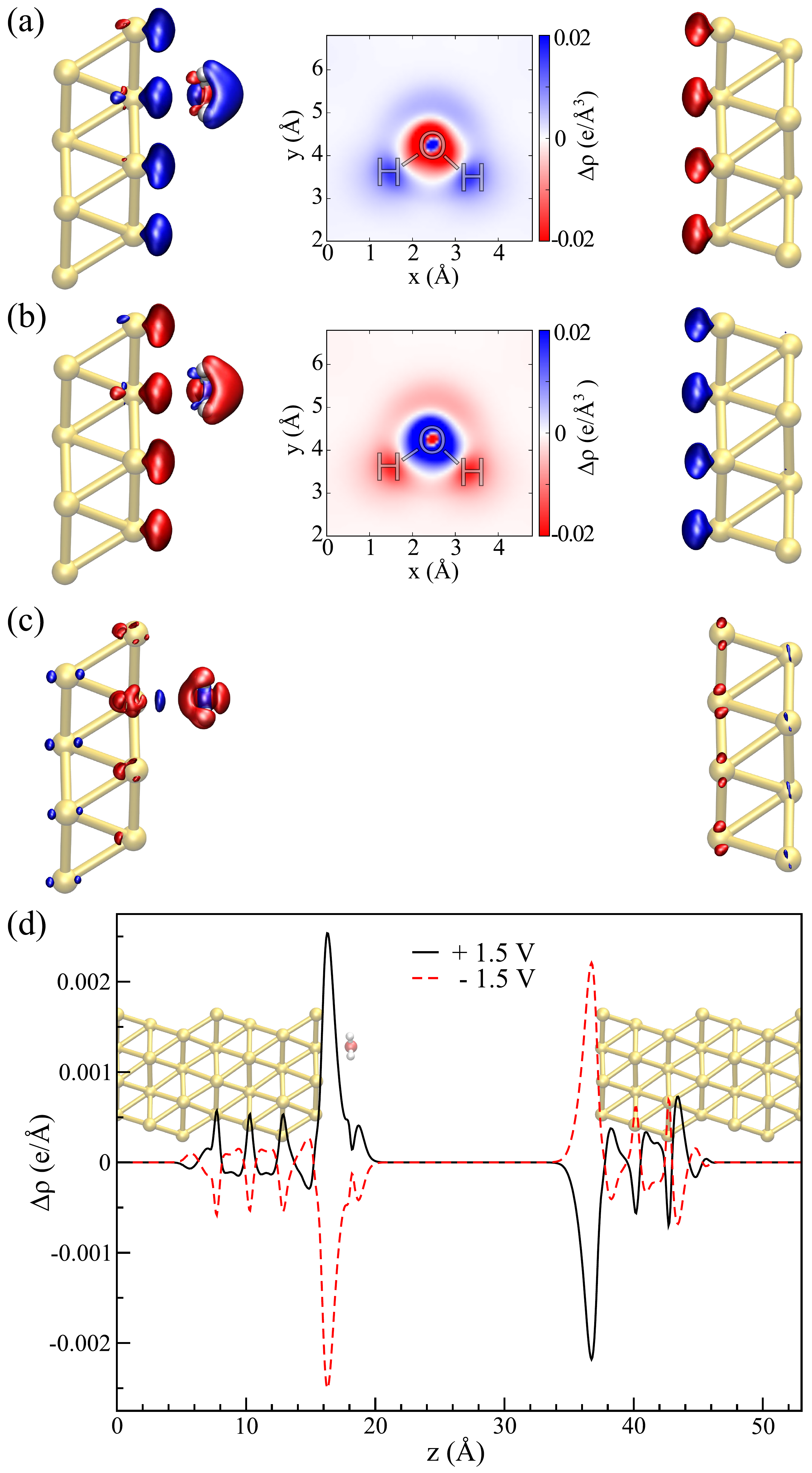}
\caption{(Color online) \emph{Flat configuration.} Relative change in charge density in comparison to the zero bias\rev{, $\Delta\rho_{V,0}=\rho_V-\rho_{0}$} for a) $V$=+1.5 V (a) and b) $V$=-1.5 V \rev{(isosurface values $\pm 8.1 \times 10^{-4}$ e/\AA$^3$). Insets show a \emph{xy}-sectional plane} taken at the water molecule \rev{center of mass}.
c) \rev{Charge density fluctuation at positive and negative applied bias, i.e. $\Sigma\Delta\rho = \Delta\rho_{1.5,0}+\Delta\rho_{-1.5,0}$ (isosurface value $\pm 5.4 \times 10^{-5}$ e/\AA$^3$).
In all cases, red (blue) indicates excess (deficiency) of electrons.
d)} Laterally averaged difference in charge density in comparison to the zero bias: in black for +1.5 V and in red for -1.5 V; the water-metal system indicates the position of the atoms.}
\label{Fig5-rho}
\end{figure}

The water-Au interaction is mostly electrostatic in nature, and there is almost no charge transfer between water and the metal in the neutral case \cite{LPedroza}, \rev{as seen from a Bader analysis \cite{bader} of the cases with and without bias (see Fig. S2 of the Supplementary Information).}
At the same time, the effect of the bias on configurations can be understood in terms of a combination of \rev{increase/decrease} in Pauli repulsion and small charging of the surface.
\figref{Fig5-rho}(a-b) show the difference in charge density for different bias compared to the zero-bias case \rev{for the ``flat'' configuration.
The corresponding insets of} \figref{Fig5-rho} indicate that most of the change in charge on the molecule is located on the oxygen.
That transfer is small, however, as seen in \rev{both the insets and} \figref{Fig5-rho}(d) \rev{which shows the change in charge density averaged over planes perpendicular to the surface.
At the same time, by calculating the fluctuations in the density differences between positive and negative bias,
\begin{eqnarray}
\Sigma\Delta\rho & = & \Delta\rho_{1.5,0}+\Delta\rho_{-1.5,0} \\
 & = & \left(\rho_{V=1.5}- \rho_{V=0}\right) + \left(\rho_{V=-1.5} - \rho_{V=0}\right) ~,
\end{eqnarray}
presented in \figref{Fig5-rho}(c) we notice, that, albeit small (the value of the isosurface is $5.4 \times 10^{-5}$ e/\AA$^3$), it is asymmetric.
Furthermore the final values of the water molecule dipole moments are similar to the ones of the isolated molecule, in the cases with and without bias, and those dipoles tend to align with the field.
The estimated magnitude of the dipole moments as a function of the applied bias are presented in the table of \figref{minima} together with the angular deviation from the dipole of an isolated molecule with the same orientation .}

\begin{figure}[ht]
\includegraphics*[width=\columnwidth,keepaspectratio]{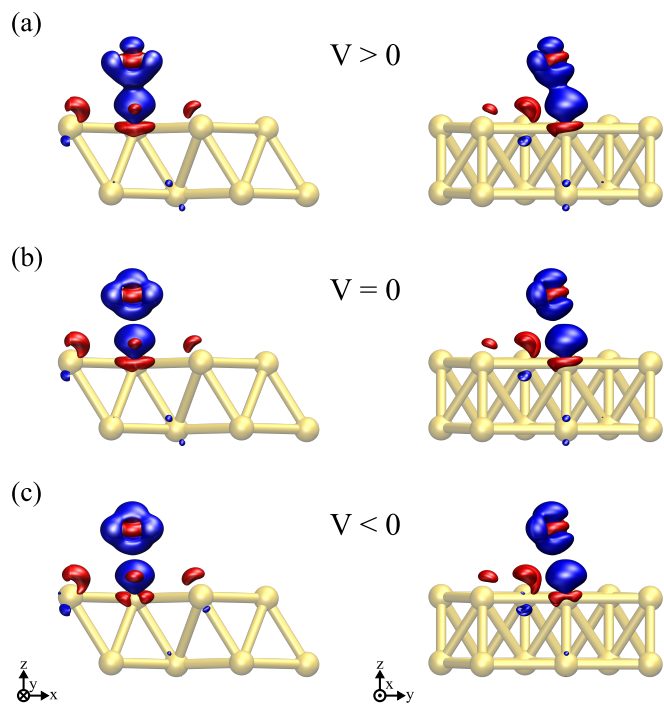}
\caption{\rev{(Color online) Difference in charge density between the Au+water system, a parallel plate capacitor and an isolated H$_2$O molecule submitted to an equivalent electric field for (a) V = 1.5 V, (b) V = 0, and (c) V = -1.5 V. An isosurface value of $\pm 8.4 \times 10^{-3}$ e/\AA$^3$ was considered in all plots, where red (blue) indicates excess (deficiency) of electrons.}}
\label{pauli}
\end{figure}

Thus, the picture that arises is the following:
for positive bias the left hand side surface becomes negatively charged and tends to repel the negatively charged oxygen.
At the same time as a small amount of charge is transferred to the oxygen the overlap between oxygen orbitals and gold surface orbitals tend to move the molecule away due to Pauli repulsion.
The opposite behavior is expected for negative bias.
\rev{This last point is evidenced in \figref{pauli} where we show the difference between our Au-water system and the charge density in an isolated capacitor with a $\pm 1.5 V$ bias applied and an isolated water molecule in an equivalent electric field.
In doing this we remove from the charge density effects that would arise solely from the electric field, focusing instead on the effects due to the water-surface interaction, namely the Pauli repulsion.
For zero bias (\figref{pauli}(b)) the signature of Pauli repulsion, namely the "pillow" density of states between molecule and metallic surface is already visible \cite{PhysRevLett.76.1138, PhysRevLett.105.086103}.
For positive bias, the interaction between molecule and surface increases and the nodes in the density disappear, an indication of decreased Pauli repulsion.
The interaction is thus more attractive.
On the other hand, for negative bias there is larger repulsion as indicated by a slightly larger gap between the "pillow" region and the charge density associated with the molecule.}

%==================================================================================================
\section{Conclusions}

In conclusion, the inclusion of electronic effects via DFT in the description of water-metal interactions is important to advance the comprehension of the local structure of water at an electrochemical interface.
In this work we showed how DFT combined with NEGF can be used to describe water-metal systems under an external bias potential.
This framework allows for a description of a truly semi-infinite metallic electrode that sets a reference chemical potential that can be controlled by applying an external bias without adding/removing additional charge to the system.
This allows for a more direct comparison with the experimental setups.
This methodology now allows to properly calculate the forces and therefore perform relaxation or dynamics of water-metal systems out of equilibrium, simulating an electrochemical cell in the sense that ionic currents could be considered now.
This can be achieved by performing a molecular dynamics simulation of the system with the bias applied.
\rev{In principle, our model allows to consider a more realistic electrochemical cell with a thicker region of water molecules, although it will be computationally more expensive.
One possibility to increase the system size in a cheaper way is to consider only the double layer region at the DFT level and the other water molecules at a molecular mechanics level (QM/MM method)\cite{QMMM1,QMMM2}.}

Here, we presented how the magnitude and sign of the bias alters the interaction of a prototype system of one water molecule on top of an Au(111) surface.
The external bias changes both the position and alignment of the molecule with the surface.
In particular, we have showed that a small positive bias leads to an unbound water molecule on a gold surface due to a combination of electrostatic effects and Pauli repulsion.
On the other hand, a negative bias increases the oxygen-metal bond, and leads to a slightly rotated water molecule, thus indicating that the introduction of an external bias has a significant influence on the microscopic structure of molecules at a metallic interface.

%%%END OF MAIN TEXT%%%

%==================================================================================================
\section{Acknowledgments}

The work was funded by DOE Early Career Award No. DE-SC0003871 and DE-FG02-09ER16052. L.S.P. and A.R.R. also acknowledge financial support from ICTP-SAIFR (FAPESP project No.\ 2011/11973-4) and the ICTP-Simons Associate Scheme.
P.B. acknowledges financial support from FP7 FET-ICT ``Planar Atomic and Molecular Scale devices'' (PAMS) project (funded by the European Commission under contract No.\ 610446), Spanish Agencia Estatal de Investigaci\'on (Grant No. MAT2016-78293-C6-4-R) and Dep.\ de Educaci\'on of the Basque Government and UPV/EHU (Grant No.\ IT-756-13).
This research used computational resources at the Center for Functional Nanomaterials, Brookhaven National Laboratory, which is supported by the US Department of Energy under Contract No.\ DE-AC02-98CH10886, and the SDumont supercomputer at the National Laboratory for Scientific Computing (LNCC/MCTI, Brazil).


%\balance


%%%REFERENCES%%%
\bibliography{auh2o} %You need to replace "rsc" on this line with the name of your .bib file
\bibliographystyle{rsc} %the RSC's .bst file

\providecommand*{\mcitethebibliography}{\thebibliography}
\csname @ifundefined\endcsname{endmcitethebibliography}
{\let\endmcitethebibliography\endthebibliography}{}

\include{SI2}

\end{document}

%% file: SI2.tex
%\documentclass[a4paper,preprint,aps,pra]{revtex4-1}
%
%%% Language and font encodings
%\usepackage[english]{babel}
%% \usepackage[utf8x]{inputenc}
%% \usepackage[T1]{fontenc}
%
%% %% Sets page size and margins
%% \usepackage[a4paper,top=3cm,bottom=2cm,left=3cm,right=3cm,marginparwidth=1.75cm]{geometry}
%
%%% Useful packages
%\usepackage{amsmath}
%\usepackage{graphicx}
%\usepackage[colorinlistoftodos]{todonotes}
%\usepackage[colorlinks=true, allcolors=blue]{hyperref}

%\vspace{0.3cm} & \vspace{0.3cm} \\

%  E-mail: l.pedroza@ufabc.edu.br
% $^{b}$~
% $^{c}$~
% $^{d}$~Department of Physics and Astronomy,  Stony Brook University, Stony Brook, New York 11794-3800, USA.
% }

%\begin{document}
\renewcommand\thefigure{S\arabic{figure}}
\renewcommand\theequation{S\arabic{equation}}

\title{Supplementary Information: Bias-dependent local structure of water molecule at a metallic interface}
%\author{Luana S. Pedroza}
%\affiliation{ICTP South American Institute for Fundamental Research,
%Instituto de F\'isica Te\'orica, Universidade Estadual Paulista, S\~ao Paulo SP 01140-070, Brazil,}
%\affiliation{ Centro de Ci\^encias Naturais e Humanas, Universidade Federal do ABC, Santo Andr\'e, S\~ao Paulo, Brazil 09210-170.}
%\author{Pedro Brandimarte}
%\affiliation{Centro de F\'{\i}sica de Materiales, 20018 Donostia-San Sebasti\'an, Basque Country, Spain,}
%\affiliation{Donostia International Physics Center, 20018 Donostia-San Sebasti\'an, Basque Country, Spain.}
%\author{Alexandre Reily Rocha}
%\affiliation{Instituto de F\'isica Te\'orica, Universidade Estadual Paulista, S\~ao Paulo, S\~ao Paulo SP 01140-070, Brazil}
%\author{M.-V. Fern\'andez-Serra}
%\affiliation{Department of Physics and Astronomy,  Stony Brook University, Stony Brook, New York 11794-3800, USA,}
%\affiliation{Institute for Advanced Computational Sciences, Stony Brook University, Stony Brook, New York 11794-3800, USA}

\maketitle

% \begin{abstract}
% Your abstract.
% \end{abstract}
\newpage
\section{Forces as a function of bias}

\begin{figure}[ht]
\includegraphics[width=\linewidth]{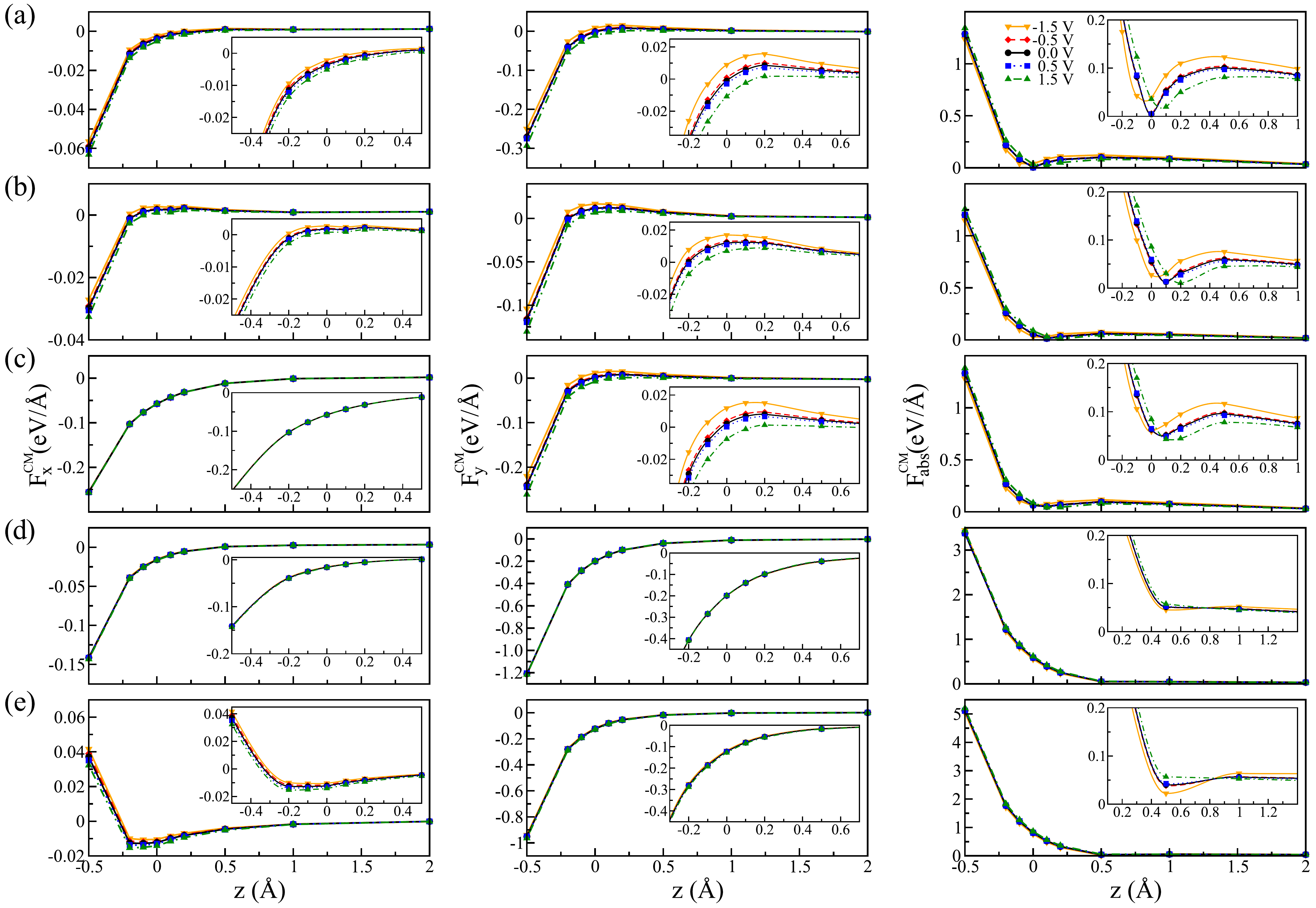}
\caption{$x$- (left panel) and $y$-components (middle panel) of the forces on center of mass of the water molecule for the different structures considered in this work (shown in Figure 2 of the main manuscript).
(right panel) Total force on the center of mass of the H$_2$O molecule.
The $z$-component is presented in the main text.}
\end{figure}

\newpage
\section{Bader Charges}

\begin{figure}[ht]
\center
\includegraphics[width=12cm]{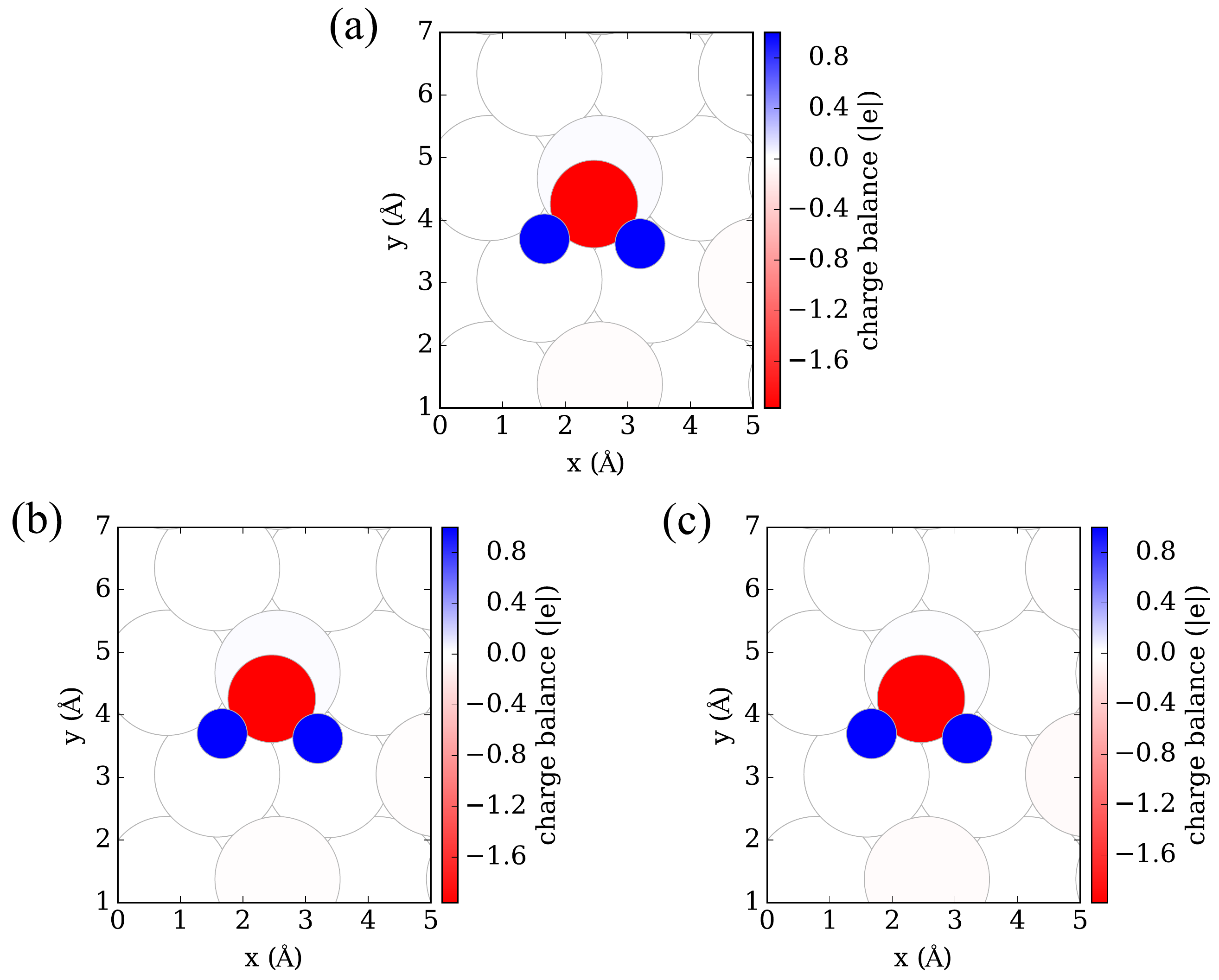}
\caption{Bader valence charges on a water molecule adsorbed on an Au surface with different bias: (a) $V$ = 0 V, (b) $V$ = -1.5 V, and (c) $V$ = 1.5 V.}
\end{figure}

\newpage
\section{Dipole moment estimation}

To analyze the effect of the electronic density redistribution on the water molecule we calculated the dipole moment $\boldsymbol{p}_i^{\mathbf{H}_2\mathbf{O}}$ for a particular arrangement $i$. In general, the dipole moment is defined in terms of the difference in charge between the converged charge density and the corresponding atomic distributions 
\begin{equation}
  \boldsymbol{p} = \int \left(\rho_{tot}{({\bf r})} - \sum_{I = 1}^{N_{atoms}} \rho_{I}{({\bf r})}\right) \left({\bf r}-\mathbf{r}_0 \right)d{\bf r}
\end{equation}
in such a way that
\begin{equation}
\int \left(\rho_{tot}{({\bf r})} - \sum_{I = 1}^{N_{atoms}} \rho_{I}{({\bf r})}\right) \sim 0 \label{neutrality} \,.
\end{equation}

As we are interested in the dipole moment of the water molecule alone, for each bias $V$, we remove the effect of an equivalent parallel plate capacitor with potential $V$,
\begin{equation}
\rho_{tot}\left(V\right) = \rho_{Au+H_2 O}\left(V\right)-\left(\rho_{Au}\left(V\right)-\sum_{I = 1}^{N_{Au}} \rho_{I}^{Au}{({\bf r})}\right) \,.
\end{equation}

All calculations are performed using $\mathbf{r}_0=\mathbf{r}_{\mathrm{CM}}$, and, in order to guarantee charge neutrality of the charge density, the limits of integration are chosen as to satisfy equation \ref{neutrality}.

%\bibliography{sample}

%\end{document}

%% file: chemScience.bbl
\begin{mcitethebibliography}{58}
\bibitem[not()]{noteEfield}
Note that the positive bias is defined when we have net negative charge on the
  surface. Therefore, this situation corresponds to the negative electric field
  in Ref.\cite{EfieldH2O}.\relax
\mciteBstWouldAddEndPunctfalse
\mciteSetBstMidEndSepPunct{\mcitedefaultmidpunct}
{}{\mcitedefaultseppunct}\relax
\EndOfBibitem
\end{mcitethebibliography}
